# Parity-breaking in single-element phases: Ferroelectric-like elemental polar metals

Hu Zhang[1,2], Bei Deng[1], Wei-Chao Wang[2], and Xing-Qiang Shi[1,3]

[1] Department of Physics, Southern University of Science and Technology, Shenzhen 518055, China

[2] Department of Electronics and Tianjin Key Laboratory of Photo-Electronic Thin Film Device and Technology, Nankai University, Tianjin 300071, China

[3] Author to whom any correspondence should be addressed.

E-mail: shixq@sustc.edu.cn

## Abstract

Polar metals based on binary and ternary compounds have been demonstrated in literature. Here, we propose a design principle for ferroelectric-like *elemental* polar metals and relate it to real materials. The design principle is that, to be an elemental polar metal, atoms should occupy at least two inequivalent Wyckoff positions in a crystal with a *polar* space group, where inversion symmetry is spontaneously broken. According to this rule, we propose the first class of potential ferroelectric-like elemental polar metals in a distorted α-La-like structure with a polar space group $P6_3mc$ in which two inequivalent Wyckoff positions 2a (0, 0, $z$) and 2b (1/3, 2/3, $z$) are occupied by group-V elements (phosphorus, arsenic, antimony, and bismuth). Analyses based on first-principles calculations indicate that the dynamically stable polar phase results from a lone pair driven polar distortion of the nonploar phase in $P6_3/mmc$ symmetry where two inequivalent Wyckoff positions 2a (0, 0, 0) and 2c (1/3, 2/3, 1/4) are occupied. This ferroelectric-like transition involves a transition from a metallic state to a semimetallic state. These predicted polar phases are metastable with respect to their corresponding ground phases. Moreover, ionic bonding characters are found due to the inequivalence in Wyckoff positions between group-V atoms. Our work opens a route to single-element parity-breaking phases.

## 1. Introduction

From the traditional viewpoint, "ferroelectricity" can only exist in insulating systems and thus all ferroelectrics have a band gap. However, ferroelectricity can exist in metallic phases. In 1960s, Anderson and Blount have already predicted the possibility of a polar metal using a phenomenological theory [1]. In 2012, Tsymbal et al. studied doped $BaTiO_3$ using combination of density functional theory calculations and phenomenological modeling [2]. They found that ferroelectric phase and conductivity can coexist. The first experimental evidence of a polar metal was reported in 2013, in which a ferroelectric-like structural transition in $LiOsO_3$ was observed [3]. More recent experimental investigations on $BaTiO_3$ using electrostatic doping also indicated the coexistence of ferroelectricity and mobile carriers [4]. New polar metals have been designed based on first-principles calculations [5,6]. These phases can also be denoted as polar metals since they are metals with polar space groups. Switching in these phases may be realized with no applied voltage [2]. Recently, the switching of polar metal domains in $Ca_3Ru_2O_7$ has been achieved experimentally [7].

Does there exist ferroelectric-like elemental polar metals (i.e. polar metals contain only atoms of one type of element)? To the best of our knowledge, no such material has been proposed to date, neither theoretically nor experimentally [8]. The current paper shows that, in fact, ferroelectric-like elemental polar metals could exist. In the following, we first propose a design principle of elemental polar phases. Then we predict realistic materials which are ferroelectric-like elemental polar metals .

## 2. Structural analyses and calculation methods

The polar phase is usually obtained from a distortion of a nonpolar high-symmetry phase. To retain the polar space group, we argue that at least two inequivalent Wyckoff positions (at least one of the sites does not exhibit inversion) need to be occupied in the unit cell of elemental polar phases (*e.g.* only containing one kind of atoms A but at inequivalent Wyckoff positions); otherwise, if occupying only equivalent Wyckoff positions, the crystal has a nonpolar space group. Our above rule can be illustrated with the following two examples. Example 1 (see also the Supplemental Material [9]), the diatomic ferroelectric GeTe has polar space group *R*3*m* (space group No. 160) [10]. Ge and Te are all in the same type of Wyckoff position 3a (0, 0, $z$) with different $z$ values in hexagonal axes. If Ge and Te are all replaced by A atoms, the crystal structure will change to nonpolar *R*-3*m* symmetry (space group No. 166) with A atoms occupying only one type of Wyckoff position 6c (0, 0, $z$) in hexagonal axes and is hence a nonpolar structure. Example 2, for ZnO in the wurtzite structure (polar space group *P*6$_3$*mc*, No. 186), two Zn atoms and two O atoms are all in the Wyckoff position 2b (1/3, 2/3, $z$) with different $z$ values. When Zn and O are all replaced by four A atoms, the crystal structure transforms into a nonpolar structure with space group *P*6/*mmm* (No. 191) in which four A atoms occupy only one type of Wyckoff positions 4h (1/3, 2/3, $z$).

For the *P*6$_3$*mc* symmetry, the Wyckoff position 2a (0, 0, z) is inequivalent with 2b (1/3, 2/3, $z$) [11]. If four A atoms occupy these two inequivalent Wyckoff positions, the crystal still has *P*6$_3$*mc* symmetry and retains polar. Thus elemental substances crystallize in this polar structure may be single-element polar metals. The corresponding nonpolar high-symmetry phase is in the *P*6$_3$/*mmc* symmetry (No. 194) in which inequivalent Wyckoff positions 2a (0, 0, 0) and 2c (1/3, 2/3, 1/4) are occupied. This nonpolar structure is related to the double hexagonal close packed (*dhcp*) structure (α-La type), which is conventionally found in rare-earth metals [12]. It should be noted that, for nonpolar phases, there exist non-centrosymmetric elemental solids if their corresponding crystallographic point groups are not polar. An example is β-tantalum (crystallographic point group D$_{2d}$), in which multiple Wyckoff positions are occupied [13].

Based on the above design principle, we then search realistic materials. To find possible elemental polar phases, we consider group-V elements (phosphorus, arsenic, antimony, and bismuth) crystallize in nonpolar *P*6$_3$/*mmc* and polar *P*6$_3$*mc* structures as mentioned above. We take antimony (Sb) as an example to demonstrate the elemental polar phases, and its underlying mechanism from first-principles calculations.

To predict physical properties of materials, the first-principles calculations was performed based on density functional theory (DFT) [14] with the local density approximation (LDA), which is usually used in the study of ferroelectrics [15], as implemented in the Vienna Ab Initio Simulation Package (VASP) [16-18]. An energy cutoff of 500 eV and a 8×8×8 Monkhorst-Pack grid [19] were used. The atomic structures were relaxed until the Hellmann-Feynman forces are less than 1 meV/Å. The phonon spectra were calculated with phonopy [20]. We also use the strongly constrained and appropriately normed meta-generalized-gradient approximation (SCAN meta-GGA) [21-23] and Heyd-Scuseria-Ernzerhof (HSE) [24]functional to check our results.

## 3. Results and discussion

We start from the nonpolar phase of Sb in the $P6_3/mmc$ symmetry. Figure 1(a) shows the crystal structure of Sb in the nonpolar $P6_3/mmc$ symmetry with four atoms per primitive cell, in which Sb atoms occupy inequivalent Wyckoff positions 2a (0, 0, 0) and 2c (1/3, 2/3, 1/4), denoted here by Sb1 (red balls) and Sb2 (blue balls), respectively. The site symmetry of $a$ ($c$) sites is $D_{3d}$ ($D_{3h}$). The optimized lattice constants are: $a$ = 4.161 Å, $c$ = 7.611 Å. This gives a lattice parameter ratio of $c/a$ = 1.829, which is much smaller than that of the perfect *dhcp* structure (about 3.2) [12]. The structure of Sb in the nonpolar $P6_3/mmc$ symmetry can be denoted as α-La-like structure. To investigate the dynamic stabilities of Sb in the nonpolar structures, we calculated the phonon spectra of Sb along high symmetry directions in the Brillouin zone, as plotted in figure 1(b). There is an unstable mode with a maximum imaginary frequency (negative values) of 78 cm$^{-1}$ around the zone center (Γ point). This unstable phonon mode is polarized along the *z*-direction with the eigenvector of [δ(Sb1) = +0.5, δ(Sb2) = −0.5], which leads to polar displacements of Sb2 relative to Sb1 as indicated by arrows in figure 1(a). The space group of the distorted structure (shown in figure 1(c)) is $P6_3mc$. This is a polar phase and can be regarded as distorted α-La-like structure. The optimized lattice constants are: $a$ = 4.151 Å, $c$ = 8.137 Å. The origin of coordinate is set on Sb1 for convenience. Sb1 (red) and Sb2 (blue) are in Wyckoff positions 2a (0, 0, 0) and 2b (1/3, 2/3, 0.1998) respectively. This polar phase is dynamically stable since no phonon modes with imaginary frequencies are found from its phonon spectra, as verified in figure 1(d). The elemental polar phase is a typical proper displacive ferroelectric-like phase with an unstable polar distortion in the nonpolar high-symmetry state.

One important parameter of ferroelectric-like is the energy barrier of the inversion symmetry-breaking phase transition from the nonpolar higher-symmetry phase to the polar phase. Figure 1(e) shows the double well for Sb. The energy difference between the polar phase and the nonpolar phase can be used to estimate the energy barrier of the realistic phase transition [15]. For Sb, it is 0.12 eV. This value is sufficiently small compared with the typical ferroelectric $PbTiO_3$ (0.2 eV) [25]. The SCAN meta-GGA functional gives a smaller energy barrier (0.07 eV) for Sb (with $a$ = 4.209 Å, $c$ = 8.163 Å). The polar phase is robust to the choice of functional.

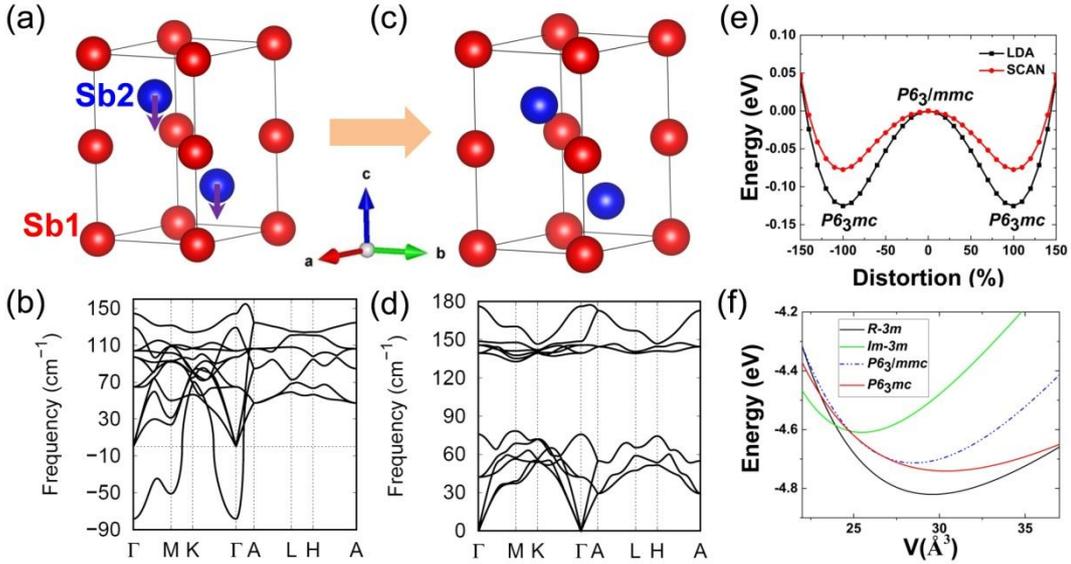

**Figure 1.** (a) The crystal structure of nonpolar Sb (space group $P6_3/mmc$) with Sb1 atoms (red balls) in Wyckoff positions 2a (0, 0, 0) and Sb2 atoms (blue balls) in 2c (1/3, 2/3, 1/4). Arrows on Sb2 indicate the polar distortion of Sb2 relate to Sb1. (b) Phonon spectrum of nonpolar Sb along the high-symmetry directions of the Brillouin zone; there is a mode with an imaginary frequency (negative values) around the Γ point. (c) The crystal structure of polar Sb (space group $P6_3mc$) with Sb1 atoms (red balls) in Wyckoff positions 2a (0, 0, 0) and Sb2 atoms (blue balls) in 2b (1/3, 2/3, 0.1998). (d) Phonon spectrum of polar Sb along the high-symmetry directions of the Brillouin zone.

(e) The double-well energy of Sb as a function of polar distortion obtained by linear interpolation between the polar and nonpolar structures calculated with LDA and SCAN meta-GGA functionals. (f) Energy-volume curves of Sb with four structures: *R-3m*, *Im-3m*, *P6$_3$/mmc*, and *P6$_3$mc*; energies per atom are given in eV.

Under ambient conditions, Sb crystallize in a structure with *R-3m* symmetry [26]. As mentioned above, this is a nonpolar phase. Sb undergoes a transition to the incommensurate host–guest structures at 8.2 and 10 GPa, and finally transforms to a body-centered cubic (bcc) structure with *Im-3m* symmetry at 28 GPa [27]. To compare the energy of these phases, we give the energy-volume curves for Sb with *R-3m*, *Im-3m*, *P6$_3$/mmc*, and *P6$_3$mc* symmetry in figure 1(f). The energy of the *P6$_3$mc* phase is 0.13 eV per atom lower than that of the *Im-3m* phase, or only 0.08 eV per atom higher than that of the *R-3m* phase. We define the formation energy (per atom) by $E_{form} = E_{atom} - (E_{bulk}/M)$, where $E_{atom}$ is the total energy of the free atom, $E_{bulk}$ is the total energy of the bulk phase and M is the number of atoms in the unit cell of the bulk phase. For Sb, we find that the formation energy of the polar *P6$_3$mc* phase predicted in our work is 0.067 eV per atom lower than that of the ground state *R-3m* phase. Advanced epitaxial growth techniques, such as molecular beam epitaxy (MBE) and Pulsed Laser Deposition (PLD) provide opportunities to obtain metastable phases. One example is Ni. The face-centered-cubic (fcc) phase of Ni is the stable phase at room temperature. The body-centered-cubic (bcc) phase of Ni is not exist in nature. However, bcc Ni has been grown on GaAs via MBE [28].

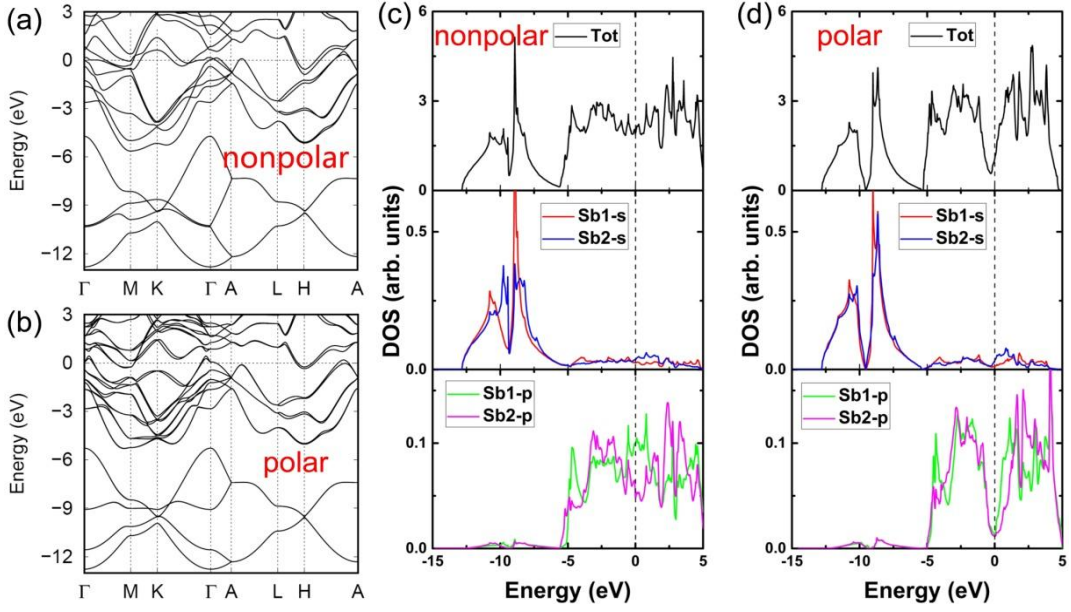

**Figure 2.** (a) Band structure of nonpolar Sb ($P6_3/mmc$) with SOC. (b) Band structure of polar Sb ($P6_3mc$) with SOC. (c) Total and partial (s and p orbitals of Sb1 and Sb2) density of states of nonpolar Sb. (d)Total and partial density of states of polar Sb. The Fermi level is set at 0 eV. These results are calculated with the LDA functional.

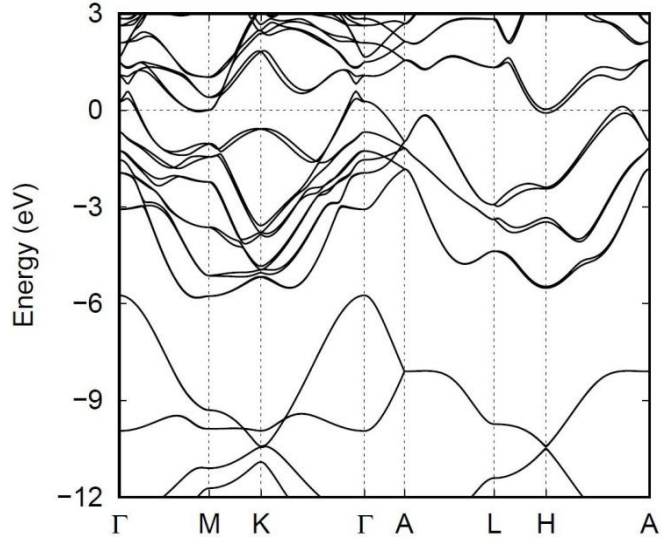

**Figure 3.** Band structure of polar Sb ($P6_3mc$) calculated with the HSE06 functional including the SOC effect. The Fermi level is set at 0 eV.

The electronic properties of the nonpolar and polar phases calculated with the LDA functional are shown in figure 2, where figure 2(a) shows the band structure of the nonpolar phase Sb calculated with spin-orbit coupling (SOC), indicating a metallic behavior of Sb. This can be confirmed again by the total density of states (DOS) shown in figure 2(c) where significant DOS at the Fermi level are found. The band structure of polar phase Sb is plotted in figure 2(b). There is a weak overlap in energy between the conduction and valence bands, which leads to the coexistence of free holes and electrons. The total DOS at the Fermi level are highly reduced in respect to the nonpolar phase, as can be found in figure 2(d). Hence, polar Sb is a semimetal. In one word, polar distortions in Sb result in a transition from metal to semimetal. The interesting new phase predicted in our work can be called the "ferroelectric-like" metals [1] or polar metals [6]. To check the LDA result, we have also calculated the band structure of polar Sb ($P6_3mc$) with the HSE06 functional shown in figure 3. This confirms that the polar Sb is a semi-metal.

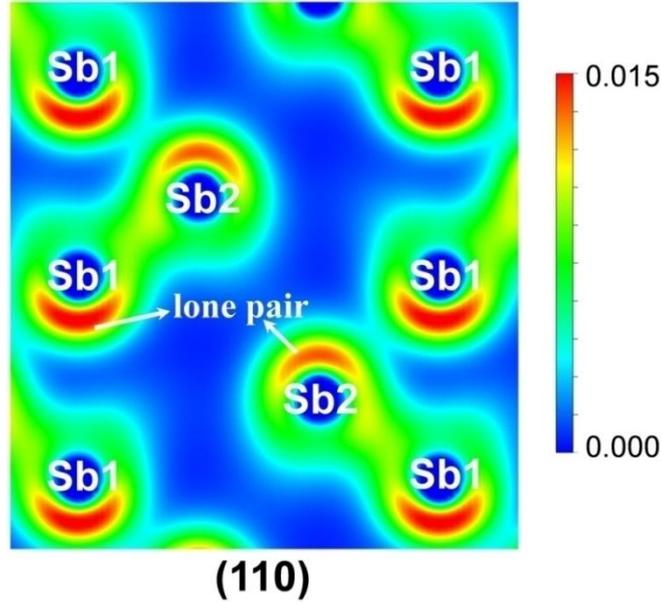

**Figure 4.** Electron densities for the states between −2 eV to the Fermi level for polar Sb ($P6_3mc$) in the (110) plane. Contour levels shown are between 0 (blue) and 0.015 e/Å$^3$ (red).

More insights into bonding characters can be found in partial electronic DOS. For nonpolar Sb, Sb1 $s$ states strongly hybridize with Sb2 $s$ states forming bonding and antibonding states located around −10 eV as shown in figure 2(c). On the other hand, the Fermi level crosses the Sb1 $p$ and Sb2 $p$ states (above −5 eV). Sb1 (Sb2) $s$ states between −5 eV and the Fermi level are formed by the antibonding of Sb1 $s$ −Sb2 $p$ (Sb2 $s$ −Sb1 $p$) combination. With the polar distortion, the filled antibonding states further mix with Sb1 (Sb2) *p* states (figure 2(d)) [29]. This leads to a highly asymmetric electron density distribution on Sb1 (Sb2) atoms, as can be confirmed from the asymmetric lobe in polar Sb shown in figure 4. Furthermore, the lone pair (asymmetric lobe) on Sb1 is more notable than that on Sb2, in line with the fact that they are in inequivalent Wyckoff positions. This is the well-known mechanism of lone pair driven structure distortions [30]. The Bader analysis [31,32] reveals that 0.05 electrons are transferred from each Sb2 atom to the Sb1 atom, which further elucidates the inequivalence between Sb1 and Sb2.

**Table 1.** Theoretical results for group-V elements (P, As, Sb, and Bi) crystallizing in an elemental form with the polar space group *P6$_3$mc*, in which two inequivalent Wyckoff positions 2a (0, 0, 0) and 2b (1/3, 2/3, *z*) are occupied. The coordinate origin is set on the atom at Wyckoff position 2a. $\Delta E$ is the energy difference between the polar state (*P6$_3$mc*) and the nonpolar state (*P6$_3$/mmc*).

|    | *a* (Å) | *c* (Å) | *z*    | $\Delta E$ (eV) |
|----|---------|---------|--------|-----------------|
| P  | 3.282   | 7.456   | 0.1600 | 0.61            |
| As | 3.633   | 7.680   | 0.1761 | 0.51            |
| Sb | 4.151   | 8.137   | 0.1998 | 0.12            |
| Bi | 4.395   | 8.158   | 0.2098 | 0.06            |

Table 1 shows the theoretical structural parameters and energy barriers for other group-V elements (P, As, and Bi) crystallized in an elemental form with the *P6$_3$mc* symmetry. They are all polar metals from electronic structure calculations. The crystal structure of the ground phase for phosphorus (P) is *cmca* [27]. The ground phases for As and Bi are the *R-3m* phase [27]. The formation energy differences between ground phases of P, As, and Bi with their predicted polar phases (*P6$_3$mc*) are 0.040, 0.061, and 0.055 eV per atom, respectively. Bader analysis reveals that 0.1 electrons are transferred between inequivalent As atoms in the distorted α-La-like structure, which shows more obvious ionic characters. We attribute the critical factor of the appearance of ionicity in single-element materials to the occupancy of inequivalent Wyckoff positions. This is different from the typical single-element materials in diamond structure (space group

*Fd*-3*m*, No. 227) with zero ionicities in chemical bonds, in which atoms (*e.g.* C, Si, Ge, and α-Sn) occupy only one type of Wyckoff positions 8*a* with coordinates (0,0,0) and (1/4,1/4,1/4).

## 4. Conclusions

In conclusion, we propose that, ferroelectric-like structure could potentially exist in single-element phases. With the help of the design principle discussed in this work, one can search other elemental polar phases (including insulating elemental ferroelectrics) with many other polar space groups. In addition, further researches could focus on other properties in single-element parity-breaking phases, *e.g.* superconductivity and electron correlation.

## Acknowledgments

This work was supported by National key research and development program (Grant No. 2016YFB0901600), the Natural Science Foundation of China (Grant Nos. 11474145, 11334003, and 21573117), the Shenzhen Fundamental Research Foundation (Grant No. JCYJ20170817105007999), and the Natural Science Foundation of Guangdong Province of China (Grant No. 2017A030310661).